\documentstyle[prl,aps,epsf,floats]{revtex}
\begin{document}
\draft
\twocolumn[\hsize\textwidth\columnwidth\hsize\csname@twocolumnfalse\endcsname

\title{Far-infrared optical conductivity gap in superconducting MgB$_2$ films}

\author{Robert~A.~Kaindl, Marc~A.~Carnahan, Joseph Orenstein, and Daniel~S.~Chemla}
\address{Department of Physics, University of California at Berkeley, and
Materials Sciences Division, E. O. Berkeley National Laboratory, Berkeley, CA 94720}
\author{Hans~M.~Christen, Hong-Ying~Zhai, Mariappan~Paranthaman, and Doug~H.~Lowndes}
\address{Solid State Division, Oak-Ridge National Laboratory, Oak Ridge, TN 37931-6056}

\date{\today}
\maketitle

\begin{abstract}
\noindent We report the first study of the optical conductivity of
MgB$_2$ covering the range of its superconducting energy gap.
Terahertz time-domain spectroscopy is utilized to determine the
complex, frequency-dependent conductivity $\sigma(\omega)$ of thin
films. The imaginary part reveals an inductive reponse due to the
emergence of the superconducting condensate. The real part
exhibits a strong depletion of oscillator strength near 5 meV
resulting from the opening of a superconducting energy gap. The
gap ratio of 2$\Delta_0/k_BT_C~\approx~1.9$ is well below the
weak-coupling value, pointing to complex behavior in this novel
superconductor. \pacs{PACS numbers: 74.70.Ad, 78.30.Er, 42.62.Fi}
\end{abstract}
]

The recent discovery of superconductivity in MgB$_2$ at~39\,K has
spawned intense research efforts, yet the nature of its remarkably
high transition temperature remains to be understood \cite{Nag01}.
While the isotope effect points to phonon-mediated
mechanisms\cite{Bud01}, its anomalous magnitude \cite{Hin01} and
distinctly different values found for the superconducting energy
gap $2\Delta_0$ \cite{Rub01,Sch01a,Tsu01,Giu01,Sza01a} are
puzzling. First-principle bandstructure calculations indicate that
the dominant hole carriers in Boron~p orbitals are split into two
distinct sets of bands with quasi-2D and 3D character
\cite{Kor01a}. Among phonon-mediated mechanisms which attempt to
explain the high $T_C$ are anisotropic two-gap scenarios where the
quasi-2D holes couple preferentially \cite{Liu01}, strong coupling
Eliashberg calculations including low and high energy modes
\cite{Man01}, and nonadiabatic superconductivity \cite{Ale01}.

The prospect of unconventional coupling or an anisotropic order
parameter motivates experiments which probe the density of states,
governed foremost by the superconducting energy gap. Surface
probes such as tunneling and photoemission found very different
values for the gap $2\Delta_0$ ranging from 4~-~14~meV in
polycrystalline MgB$_2$ \cite{Rub01,Sch01a,Tsu01,Giu01,Sza01a}.
Possible reasons for this large variation might include proximity
effects or a distribution of surface composition that remains
hidden in the bulk T$_C$ \cite{Sch01a,Tsu01}. The results could
also originate from a multigap mechanism, and indeed recent point
contact experiments found regions where two distinct conductance
peaks of different magnitude appear ($2\Delta_0 = 6$ and 14 meV)
\cite{Giu01,Sza01a}.

Optical probes offer important advantages since they penetrate
deeply inside the bulk, where they probe the low-energy
excitations including the superconducting energy gap
~\cite{Pal68}. In MgB$_2$, a first measurement of far-infrared
reflectivity in granular samples showed an increase below
$\hbar\omega\lesssim9$~meV \cite{Gor01}, while the transmission
through films also increases in that range \cite{Jun01}. While
this could stem from a gap in the real part of the optical
conductivity, the necessary Kramers-Kronig transformations mandate
measurements over a much broader spectral range and with high
precision. Thus, direct measurements of the conductivity are
crucial. First results at low photon energies (0.5~-~3.7~meV),
however, found no characteristic signature of a BCS gap in that
range \cite{Pro01}.

In this Letter, we present the first measurement of the
far-infrared conductivity of MgB$_2$ in a broad frequency range
that spans excitations across its superconducting gap. The complex
conductivity $\sigma(\omega)$, measured in transmission, reveals
information about the fundamental low-energy excitations. Below
T$_C$ the real part of the conductivity is dominated by the
depletion of oscillator strength due to the opening of a
superconducting gap, while the imaginary part displays the
inductive response of the condensate. The normal state scattering
rate exceeds the gap energy. We find a gap size
of~$2\Delta_0\approx$~5~meV, which is only half the value expected
for an isotropic BCS gap.

Highly crystalline, superconducting films of MgB$_2$ were grown in
a two-step process. Deposition of B precursor films via
electron-beam evaporation was followed by ex-situ post-annealing
at 890~$^\circ$C in the presence of bulk MgB$_2$ and Mg vapor
\cite{Par01}. For the optical measurements, we investigate films
of 100 and 200~nm thickness on Al$_2$O$_3$ substrates, with
corresponding T$_C$'s (zero resistance) of~30.5~K~and 34~K~
(inset, Fig.~1) \cite{comm0}. Scanning electron microscopy shows
dense films with surface roughness below 5~nm for the 100~nm film,
and a grain-like morphology similar to Ref. \cite{Par01} was found
for the thicker film.

In our experiment terahertz time-domain spectroscopy is used to
determine the {\it complex} transmission function in the spectral
range of interest without the need for Kramers-Kronig
transformation \cite{Nus98}. Femtosecond near-infrared pulses from
a 250~kHz Ti:sapphire amplifier are focussed onto 0.5~mm~thick
$\langle110\rangle$ oriented ZnTe crystals to generate and detect
far-infrared pulses. This broadband far-infrared radiation
(2-11~meV photon energy) is transmitted through the MgB$_2$ film
mounted at normal incidence in a continuous-flow He cryostat
equipped with Mylar windows.

\begin{figure}[b!]
\epsfxsize=3.0in \centerline{\epsffile{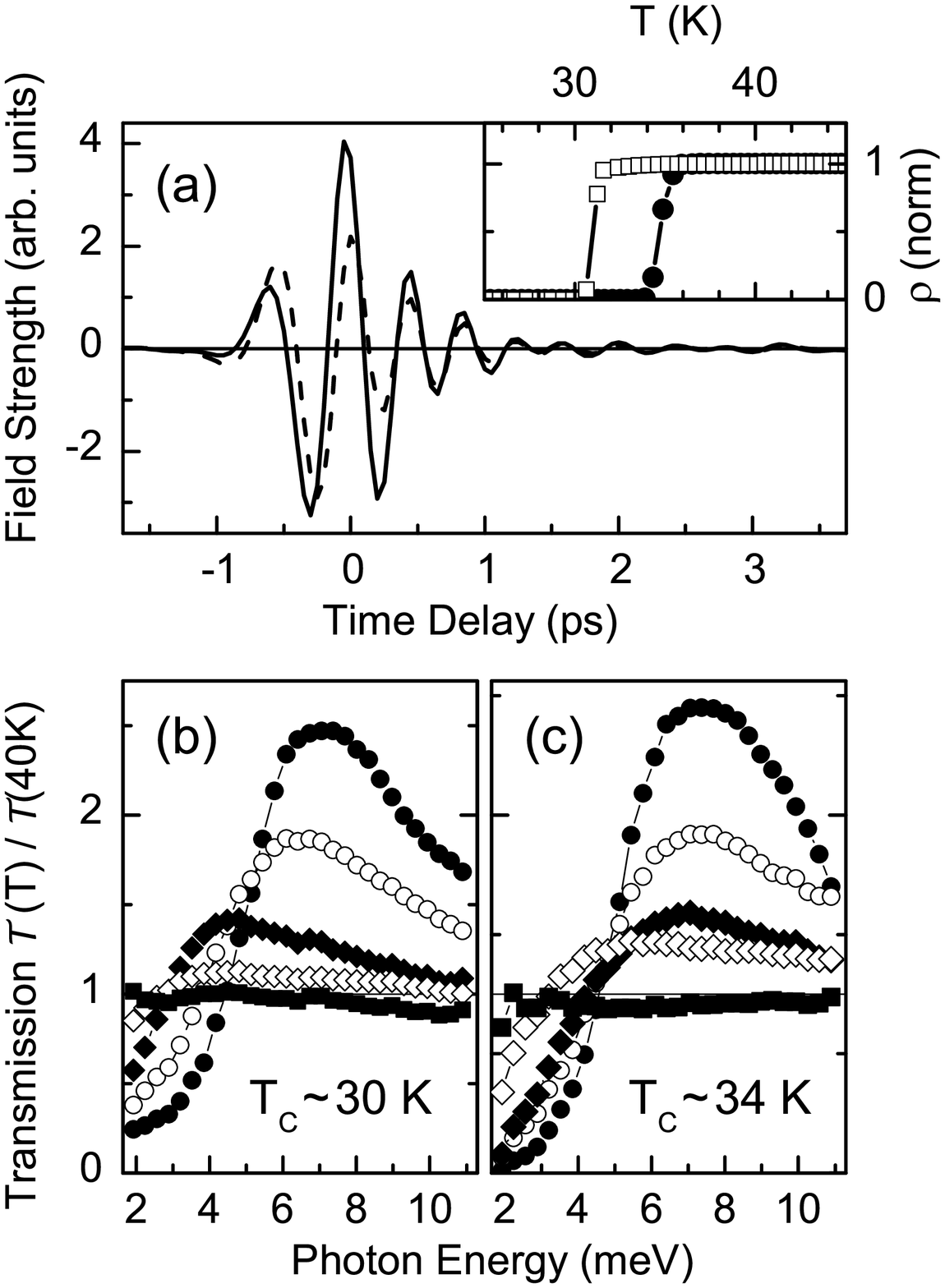}}
\vspace{0.2cm} \caption{(a) Electric field transients transmitted
through the 100\,nm MgB$_2$ film at T\,=\,6\,K (solid line) and
40\,K (dashed line). Inset: resistance of the 200\,nm (dots) and
100\,nm film (open squares) corresponding to
$\rho(40\,K)\,\approx\,10$ and $100\,\mu\Omega$cm, respectively.
(b) Transmission ${\cal{T}}$ normalized to ${\cal{T}}$(40\,K) as
obtained from the transients for the 100 nm thick film at
T\,=\,6\,K (dots), 20\,K (open circles), 27\,K (solid diamonds),
30\,K (open diamonds), and 33\,K (solid squares). (c) results for
the 200\,nm thick film at T\,=\,6\,K (dots), 20\,K (open circles),
25\,K (solid diamonds), 30\,K (open diamonds), and 36\,K (solid
squares).} \label{fig1}
\end{figure}

Figure 1(a) shows the measured time-dependent electric field
$E(t)$ of far-infrared pulses transmitted through the MgB$_2$ film
in the superconducting and normal states. Below T$_C$ (solid
line), the pulse exhibits an increase of its field amplitude along
with an apparent phase shift. This reshaping is linked to the
changes of the frequency-dependent conductivity in the
superconducting state. Fourier transformation of the incident and
transmitted fields, $E_i(t)$ and $E_t(t)$, yields spectral
information via the complex transmission coefficient $t(\omega) =
E_t(\omega)/E_i(\omega)$. The power transmission spectrum
${{\cal{T}}(\omega)} = |t(\omega)|^2$, normalized to its 40~K
normal state value, is shown in Figs.~1(b) and (c) for the 100\,nm
and 200\,nm thick samples, respectively. Above T$_C$ (solid
squares), the transmission remains unchanged, but below~T$_C$ a
transmission increase is observed in a broad spectral range. Its
peak shifts to higher photon energies with decreasing temperature.
At the lowest temperature (6K, dots), a more than twofold increase
is found for photon energies around 7~meV, concurrent with a
strong decrease below~4.5~meV. The response of the two samples is
similar and, most notably, the spectral features occur at an
identical position.

A more detailed understanding can be derived from the frequency
dependent complex conductivity $\sigma(\omega)\,=\,
\sigma_1(\omega)\,+\,i\sigma_2(\omega)$ of the film, which is
obtained from the measured transmission function at normal
incidence in the thin-film limit via $\sigma =
((1+n_s)~t_{f}/t_{s}\,-\,n_s\,-\,1)/(Z_0 d)$ ~\cite{Nus98}. Here,
$Z_0$ is the impedance of free space, d~the film thickness, $n_s$
the substrate refractive index, and $t_{f}$ and $t_{s}$ the
complex transmission coefficients of film + substrate, and
substrate, respectively. The thin-film approximation ($n_{\em
film} \omega d / c \ll 1$) is fulfilled well for the 100~nm film
(transmission coefficient $|t(\omega)| \approx~0.1$), but the
200~nm film is optically too thick and thus was not evaluated
quantitatively in this manner.

Absolute values of the normal state conductivity are shown in the
inset of Fig. 2 for both the real part (circles) and imaginary
part (squares). In the simplest case, the results can be compared
to a single-component Drude conductivity $\sigma(\omega) =
\epsilon_0 \omega_{pl}^2 / (\tau^{-1} - i\omega)$. The parameters
are constrained by fitting the real and imaginary part
simultaneously (solid lines), which yields a plasma frequency
$\omega_{pl} = 12000~$cm$^{-1}$ (1.5~eV) and scattering rate
$\tau^{-1} = 300~$cm$^{-1}$ (37~meV). Even for a strong coupling
gap of $\approx$\,10\,meV this corresponds to the so-called dirty
limit $\hbar / \tau \gg 2\Delta$ in which absorption should set in
for photon energies exceeding the gap energy $2\Delta$ since
elastic scattering enables momentum conservation to the final
state.

The ratio of the experimentally determined real part of the
conductivity $\sigma_1(\omega)$ to its normal state value is
plotted in Fig. 2 for different temperatures. As expected, the
changes are small above T$_C$ (diamonds), yielding a ratio close
to one. As the temperature falls below T$_C$ a strong depletion
of $\sigma_1$ is observed: the conductivity is smallest around 4.5
meV and then increases monotonically at higher photon energies.
The steepest slope is obtained at the lowest temperature (6\,K,
dots) revealing a strong absorption onset around $\approx 5 meV$.
Results for the corresponding imaginary part of the conductivity
$\sigma_2(\omega)$ are given in Fig.\,3(a) [symbols]. In the
superconducting state $\sigma_2(\omega)$ shows the typical shape
indicative of the supercurrent's high-frequency electromagnetic
reponse, falling off strongly with photon energy and decreasing
with temperature.

\begin{figure}[t!]
\epsfxsize=3.2in \centerline{\epsffile{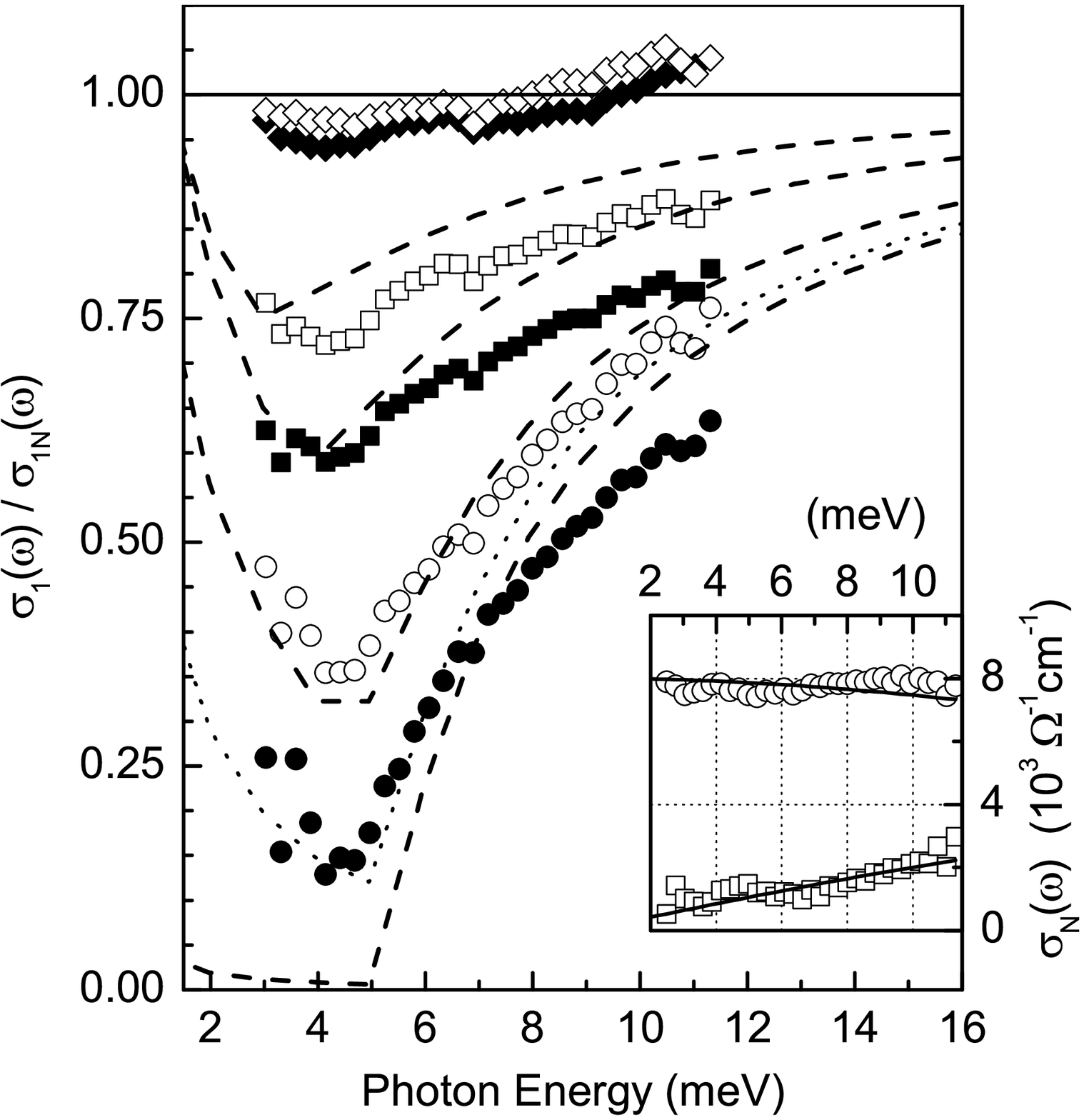}}
\vspace{0.2cm} \caption{Real part of conductivity
$\sigma_1(\omega)$ for the 100\,nm film normalized to its normal
state value $\sigma_{1N}$(40\,K) for T\,=\,6\,K (dots), 17.5\,K
(open circles), 24\,K (solid squares), 27\,K (open squares), 30\,K
(solid diamonds), 50\,K (open diamonds). Mattis-Bardeen
calculations for $2\Delta_0\,=\,5$\,meV, T$_C$\,=\,30 K are shown
for (dashed and dotted lines, bottom to top) 6\,K, 12\,K, 17.5\,K,
24\,K, and 27\,K. Inset: real (circles) and imaginary part
(squares) of normal state conductivity along with a Drude
calculation (lines) explained in the text.} \label{fig2}
\end{figure}

We now discuss the spectral shape and temperature dependence of
the observed far-infrared conductivity. For comparison,
calculations were performed using the theory of Mattis and Bardeen
for an isotropic s-wave gap~\cite{Mat58}. In view of the lower
energy threshold of the observed absorption onset, an artificially
small gap value of $2\Delta_0~=~5$~meV was chosen. The results of
such calculations at different temperatures are shown in Fig. 2
(dashed lines) where the gap $\Delta(T)$ was assumed to follow the
usual BCS temperature dependence. The calculated absorption sets
in above $2\Delta_0$ and converges to the normal state
conductivity at higher photon energies. In addition, a
low-frequency Drude-like component due to thermally activated
normal carriers gains weight with increasing temperature. We
emphasize that the experimental data follow this overall trend in
frequency and temperature dependence. A slower rise with photon
energy is observed in the experiment, however, and some
low-frequency residual conductivity remains at the lowest
temperature which could also result from a slightly larger film
temperature (dotted line, Fig. 2). Most strikingly, the value of
the absorption onset is almost a factor of two smaller than even
the value $2\Delta_0= 3.5\,k_B\,T_C\,\approx\,9 meV$ expected in a
weak-coupling scenario.

\begin{figure}[b!]
\vspace{0.3cm} \epsfxsize=3.3in
\centerline{\epsffile{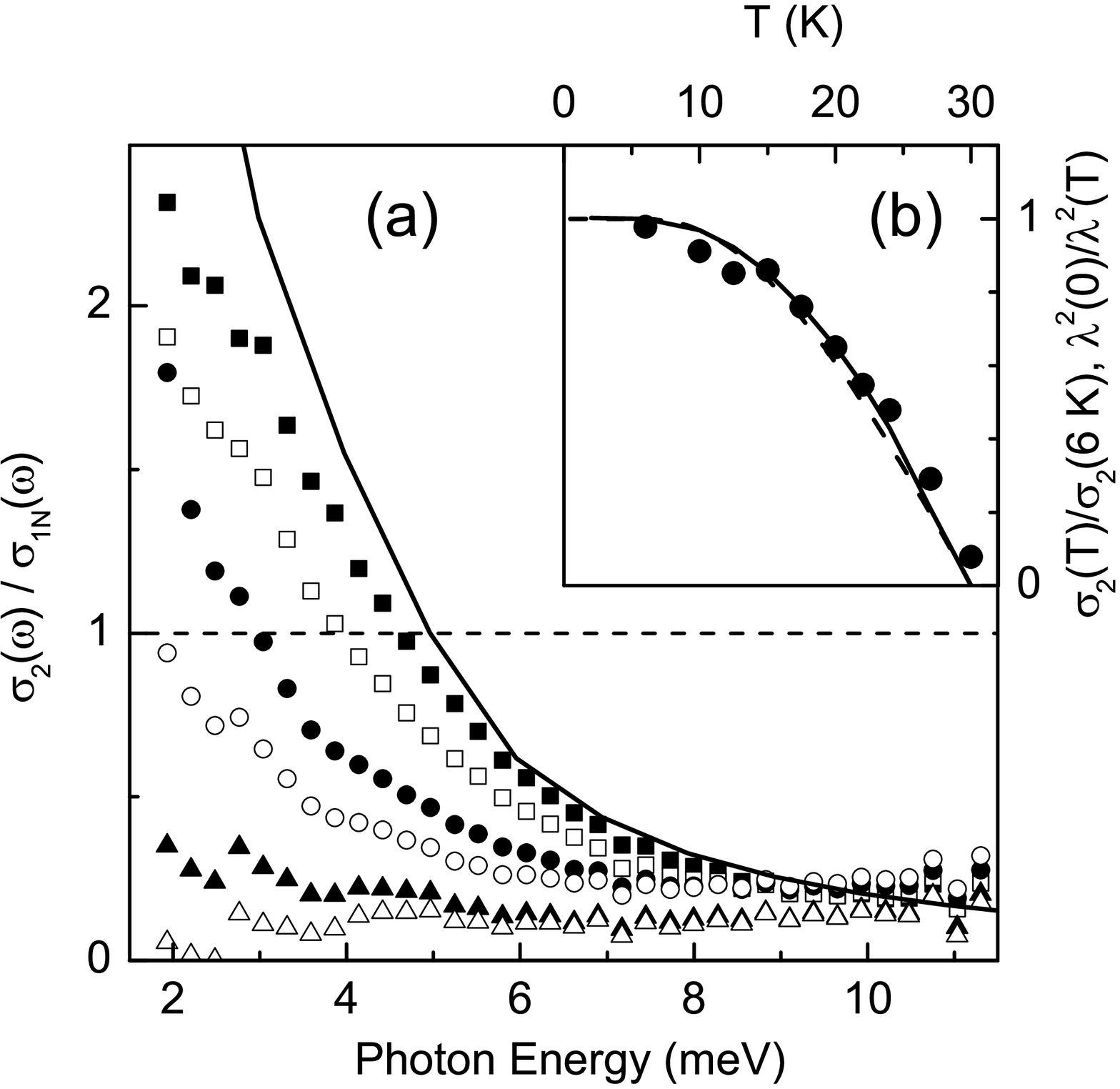}} \vspace{0.2cm} \caption{(a)
Imaginary part of the conductivity, $\sigma_2(\omega)$, normalized
to the normal state value $\sigma_{1N}$ at T~=~40 K. Results are
shown for T~=~6~K (solid squares), 17.5~K (open squares), 24~K
(dots), 27~K (open circles), 30~K (solid triangles), 33~K (open
triangles). Solid line: Mattis-Bardeen calculation for T\,=\,6\,K
(parameters see Fig. 2). (b) Normalized temperature dependence of
$\sigma_2$(3\,meV) [solid circles], compared to the Mattis-Bardeen
result at this frequency (solid line) and the BCS penetration
depth (dashed line).} \label{fig3}
\end{figure}

The calculated imaginary part of the conductivity shown in Fig.
3(a) (solid line) for the lowest temperature of 6~K comes close to
the experimentally observed frequency dependence. Whereas Pronin
et al. found $\sigma_2(\omega)\sim\omega^{-1}$ at much lower
frequencies \cite{Pro01}, we observe here that $\sigma_2(\omega)
\sim \omega^{-2}$ yields a more faithful representation of the
data for larger photon energies $\hbar\omega\gtrsim 5~meV$. This
is in agreement with the Mattis-Bardeen calculation, and is
microscopically explained by the onset of dissipative excitations
across the gap. The penetration depth $\lambda$ can be obtained
from the optical conductivity at frequencies sufficiently below
the gap, where $\lambda^{-2}(T) = \mu_0 \omega
\cdot~\sigma_2(\omega,T)$. We estimate
$\lambda(0)~\approx~3000$\,\AA~from our data at the lowest
available temperature and photon energy. Further insight is
obtained from the temperature dependence of $\sigma_2$, which at
low frequencies should follow the square of the penetration depth.
Figure~3(b) displays the measured temperature dependent
conductivity at 3\,meV~(dots). Within the experimental accuracy,
and in contrast to Ref. \cite{Pro01}, satisfactory agreement is
found with the Mattis-Bardeen calculation at the given frequency
(solid line). The data thus indicate that despite the lowered gap
size the penetration depth follows the BCS behaviour quite well.

One particularly intriguing result of this study is the
persistently small size of the conductivity gap as compared to
the transition temperature. Here, we find a ratio of only
$2\Delta_0/k_B T_C = 1.9$, which is far below the usual weak or
strong coupling BCS values. Our experiment shows that the
dramatic transmission changes observed in this spectral range
(Fig. 1b) are ultimately linked to the emergence of this
conductivity gap as in conventional metals \cite{Pal68}. The {\it
identical} spectral position of such transmission changes for both
samples investigated here, as well that from a different study
\cite{Jun01}, point to a more robust nature of this observation.

We emphasize that this reduced ratio cannot be due to a sample
inhomogeneity in which percolative paths through regions with
increased gap would yield the transport T$_C$, whereas the bulk
fraction of the sample would become superconducting at a much
lower temperature linked to the smaller gap via the usual BCS
ratio. This scenario is not possible since our optically measured
imaginary part of the conductivity, which probes the complete
volume of the film, shows directly that the major fraction of the
condensate persists up to the large transport-derived~T$_C$.

Different mechanisms might explain the observed small gap. First,
it is interesting to ask to what extent the two-gap scenario might
apply \cite{Liu01}, where our observed conductivity gap would
correspond to the smaller order parameter, whereas the larger one
lies outside our optically accessible range. We observe that
$\sigma_1(\omega)$ rises slower with photon energy than predicted
by the isotropic calculations, which might support this
contention. Yet, we also note that extended strong-coupling
calculations including a modified phonon spectral density allow a
smaller gap ratio to emerge even in an isotropic Eliashberg
formalism \cite{Man01}. However, a calculation of the optical
conductivity in these models for MgB$_2$ has not been carried out
so far.

Anderson's theory of superconductors in the dirty-limit shows that
for an isotropic order parameter neither the gap size nor $T_C$
are affected by the nonmagnetic impurity scattering events that
prevail in our films \cite{And59,comment2}. Nevertheless, a more
general anisotropic gap could be averaged out in this case,
evoking a gap ratio that approaches the usual weak or strong
coupling values. However, the two-gap state should persist for
predominantly {\it intra}-band nonmagnetic scattering
\cite{Gol97}, e.g. for small-angle impurity scattering which lacks
the momentum to scatter holes between the {quasi-2D} and 3D Fermi
surfaces well separated in momentum space. Comprehensive
theoretical calculations of the optical and other fundamental
physical properties are imperative to achieve a full understanding
of the low-energy excitations in MgB$_2$.

In summary, we have studied the far-infrared conductivity of thin
MgB$_2$ films using terahertz time-domain spectroscopy in a broad
spectral range. The complex conductivity exhibits the
characteristic electrodynamic response of a dirty-limit metal in
the normal state. An inductive response in the imaginary part
appears below $T_C$ due to the emergence of the superconducting
condensate. A strong depletion of the real part of the
conductivity corresponds to the opening of the superconducting
gap, yet its energy threshold 2$\Delta~\approx$~5~meV is only half
that expected in an isotropic, weak-coupling theory.

This work was supported by the US Department of Energy, Office of
Science, under contracts DE-AC03-76SF00098 and DE-AC05-00OR22725.
R.A.K. gratefully acknowledges financial support from the Deutsche
Forschungsgemeinschaft.


\begin{references}
\bibitem{comment2} We rule out magnetic impurities since $(i)$
energy-dispersive x-ray analysis shows no detectable traces of
such elements $(ii)$ expected strong variations of the spectral
gap position with $T_C$ remain unobserved (cf. Fig.\,1)
\end{references}
\end{document}